# How Far Do Capability Cues Travel? Anthropomorphism and Differentiated Trust in a Platform-Embedded AI Assistant

CHENCHEN MAO, Lehigh University, United States
HANJING SHI, Lehigh University, United States
HAIYAN JIA, Lehigh University, United States
DOMINIC DIFRANZO, Lehigh University, United States

Visible AI capabilities need not translate into broader judgments of trustworthiness. In a randomized $2 \times 2$ experiment with 270 U.S.-based Reddit users, an embedded assistant displayed one or three functions, with or without a brief rationale. Displaying three functions increased perceived multifunctionality; no other randomized main effect survived correction across the six outcomes. Rationale availability did not reliably increase perceived intelligence. Exploratory analysis indicated stronger uptake of the functional display at higher objective AI literacy. Among concurrently measured judgments, perceived multifunctionality was associated with perceived intelligence, which was associated with anthropomorphism and all three trust dimensions. After accounting for perceived intelligence, anthropomorphism was positively associated with benevolence, but not reliably with integrity or ability. These findings separate interface effects from relationships among users' perceptions and show why ability, integrity, and benevolence should be evaluated separately.

## 1 Introduction

General-purpose AI (GPAI) refers to AI systems that can perform a wide variety of tasks across domains rather than being limited to a single narrow function [47]. A prominent example is ChatGPT. In recent years, such systems have improved rapidly in raw performance, in some cases surpassing human baseline performance on established benchmarks [34]. Contemporary large language model (LLM)-based systems have also made two capabilities especially visible to users. They can perform multiple distinct tasks, including summarization, classification, and sentiment analysis [1], and they can generate rationales that justify or clarify their outputs [17]. Multifunctionality and rationale generation are therefore two salient, interface-relevant capabilities of contemporary AI assistants.

Alongside these developments, AI assistants increasingly appear inside the platforms where people are already reading and responding to content [18, 36, 45]. On a social-media page, the assistant appears alongside the post as a supporting feature and may be judged from a brief presentation rather than extended use. The value of this assistance depends not only on what the system provides but also on whether users are willing to consider and rely on it [24, 26]. Understanding perceptions of the assistant—particularly its trustworthiness—is therefore important for explaining how its assistance may be received and used. We therefore focus on whether and how the two features introduced above—multifunctionality and explanation—shape users' evaluations of the embedded assistant.

Trust rests on several judgments [35]. A user may regard an assistant as capable while remaining uncertain whether it follows acceptable principles or acts with the user's interests in mind. A single overall trust score can obscure these differences. Understanding trust in an embedded assistant therefore requires distinguishing these judgments and examining how they relate to users' perceptions of the assistant.

Authors' Contact Information: Chenchen Mao, chm321@lehigh.edu, Lehigh University, Bethlehem, Pennsylvania, United States; Hanjing Shi, hasa23@lehigh.edu, Lehigh University, Bethlehem, Pennsylvania, United States; Haiyan Jia, haj616@lehigh.edu, Lehigh University, Bethlehem, Pennsylvania, United States; Dominic DiFranzo, djd219@lehigh.edu, Lehigh University, Bethlehem, Pennsylvania, United States.

We organize these perceptions and trust judgments into three layers. *Capability perception* includes perceived multifunctionality—the breadth users attribute to the assistant—and the broader judgment of perceived intelligence. *Social attribution* concerns anthropomorphism, the attribution of personlike characteristics or intentions [8]. *Differentiated trust* includes ability, integrity, and benevolence: whether the assistant seems competent, principled, and attentive to the user's interests [35]. Perceived intelligence and anthropomorphism are related but distinct judgments [23], while prior research shows that social cues can influence different dimensions of trust unequally [22].

At the capability-perception layer, we examine displayed functional breadth and explanation availability as two interface cues. For multifunctionality, we distinguish *assigned* breadth, what the interface displays, from *perceived* breadth, what users attribute to the assistant. This distinction follows the view that visible interface features provide cues from which users infer a system's capabilities [9, 46]. We therefore examine whether a broader functional display increases perceived breadth and whether perceived breadth is associated with perceived intelligence [25, 31]. For explanation, we examine whether providing a rationale increases perceived intelligence. Research on agent transparency and perceived explainability provides related evidence linking explanation-related features to cognitive evaluations of AI systems [53, 54]. However, providing a rationale does not guarantee that users will find it useful, because its value may depend on whether it addresses their questions [27, 37]. We therefore examine whether participants shown a rationale report higher perceived intelligence, on average, than those shown outputs alone.

Two questions follow: how far do these cues travel into capability perception, and how are perceived intelligence and anthropomorphism associated with the three dimensions of trust? Section 3 develops these questions into four hypotheses and one research question. Motivated by research on AI literacy and how users interpret and respond to AI cues [7, 10, 28, 51], we conducted an exploratory screen of objective AI literacy as a potential moderator across the relationships in the hypothesized model.

To answer these questions, we conducted a randomized $2 \times 2$ between-participants experiment with 270 U.S.-based Reddit users. Using the *Outer Limits* browser extension [32], we embedded an assistant in a locally rendered Old Reddit page. We manipulated displayed functional breadth by presenting either one function or three distinct functions, and explanation availability by providing or omitting a brief rationale for the assistant's outputs. The post, comments, and engagement indicators were held constant across conditions.

Displaying three functions increased perceived multifunctionality. This effect was the only randomized main effect that survived correction across the focal outcomes. The total effect of assigned multifunctionality on perceived intelligence did not survive correction. Providing an explanatory rationale likewise did not significantly increase perceived intelligence. An exploratory analysis indicated that uptake of the functional display was stronger among participants with higher objective AI literacy. The remaining evidence is associational: perceived multifunctionality was associated with perceived intelligence, which was associated with anthropomorphism and all three trust dimensions. After accounting for perceived intelligence, anthropomorphism was additionally associated with benevolence, but not with integrity or ability.

This paper makes two contributions. First, it shows the limits of using visible functionality and explanatory rationales to communicate capability. Displaying multiple functions increased perceived functional breadth, but neither this display nor the provision of a rationale produced a statistically reliable increase in perceived intelligence. Second, it connects capability perception to social attribution while showing that the two judgments carry different trust profiles. Perceived intelligence was positively associated with anthropomorphism and with ability, integrity, and benevolence. After accounting for perceived intelligence, anthropomorphism was associated specifically with benevolence. These findings matter because feature recognition alone cannot establish the broader success of an interface intervention, and a global trust score can conceal whether users regard the assistant as capable, principled, or attentive to their interests. Evaluations of embedded AI should measure the specific judgment a feature is intended to change and the particular dimension of trust at stake.

## 2 Related Work

### 2.1 Reading Capability from Interface Cues

Research on machine agency treats interface features as cues from which users infer what an AI system can do and how independently it can act. Sundar [46] distinguishes a *cue route*, through which visible interface signals shape perceptions of machine agency, from an *action route*, through which judgments develop through active engagement, volitional control, and interaction with the system. Research on algorithmic news feeds similarly shows that users construct understandings of otherwise opaque systems from the content and controls made visible to them [9]. The setting examined here concerns the cue route: users encounter a bounded presentation of an assistant rather than learning about it through repeated interaction or changes in its performance over time.

What an interface displays and what a user concludes are not necessarily the same. Studies of conversational agents document gaps between expected and experienced capabilities [29], while controlled experiments show that stated and observed system accuracy can have different consequences for trust [52]. Displaying several functions may therefore communicate a broad functional repertoire without ensuring that every user perceives the assistant as multifunctional. This distinction separates the capability breadth made visible by an interface from the capability breadth users attribute to the system.

Explanation presents a different type of potential capability cue. Beyond showing what an assistant can do, a rationale can make the apparent basis for an output visible and thereby provide users with additional information for judging how the system reached its response. Prior research provides evidence that explanation-related transparency can inform cognitive evaluations of an AI system. Greater agent transparency increased perceived intelligence in a driving-simulator study [53], while perceived explainability has been positively associated with perceived intelligence and other cognitive evaluations of automated systems [54]. Together, these findings suggest that a visible rationale may be interpreted as evidence of cognitive capability.

At the same time, the presence of an explanation does not guarantee that it will function as such a cue. Explanation is an audience-sensitive communicative process [37], and its value may depend on whether it addresses the questions users want answered [27]. Explanation availability and explanation usefulness are therefore not equivalent. A rationale may provide a basis for judging the assistant's intelligence, but whether users make that inference remains an empirical question.

Users may also differ in how they interpret the same interface cues. AI literacy includes competencies for recognizing, understanding, critically evaluating, and using AI technologies across contexts [28]. Prior AI knowledge can affect what users take from an explanation [7], while training can alter anthropomorphic language without producing corresponding changes in self-reported anthropomorphism or trust [51]. This literature suggests that AI knowledge may shape cue interpretation, but it does not establish a consistent direction or identify a particular relationship at which such differences should emerge.

### 2.2 Perceived Intelligence and Anthropomorphism

The Computers Are Social Actors literature suggests that people may apply social expectations to computers with relatively little deliberation [41]. However, such responses are not inevitable. A recent direct replication using familiar desktop computers found narrower evidence for CASA effects and suggested that technological novelty may influence when they occur [16]. Contemporary AI assistants provide visible capability cues through generated language, task performance, and rationales. The resulting question is therefore not only whether users respond socially to these systems, but which capability perceptions contribute to seeing them in personlike terms.

Perceived intelligence is one possible basis for anthropomorphic judgments. Research on conversational agents shows that users evaluate an agent's intelligence alongside socially oriented characteristics such as approachability, individuality, and its apparent attitude toward the user [23]. These judgments are related but conceptually distinct. Perceived intelligence concerns whether a system appears knowledgeable and cognitively capable, whereas anthropomorphism concerns the attribution of personlike characteristics, motivations, or intentions. The three-factor theory of anthropomorphism proposes that knowledge elicited by an agent's observable characteristics and behavior can provide material for interpreting it in humanlike terms [8]. Prior work has accordingly examined perceived intelligence and anthropomorphism across personal intelligent agents, conversational agents, and chatbot interfaces [30, 38–40]. More specifically, positive associations between these judgments have been reported in studies of personal intelligent agents, a virtual teaching assistant, and a plain-text AI system [33, 39, 49]. This literature supports treating cognitive competence and personlike attribution as distinct but potentially related judgments.

### 2.3 Trust as a Differentiated Judgment

Trust in AI is often analyzed as a global evaluation. A meta-analysis of 65 articles found that system performance and anthropomorphism were among the factors associated with trust in AI [21]. Other traditions treat trustworthiness as multidimensional. The ability–integrity–benevolence framework distinguishes whether a trustee appears competent, principled, and oriented toward the trustor's interests [35]; this framework has also been applied to online recommendation agents [50]. Research on personal intelligent agents further suggests that cognitive and social perceptions may correspond to different forms of trust: perceived intelligence has been linked to cognitive trust, whereas anthropomorphism has been linked to emotion-based trust [38].

Prior findings nevertheless do not establish a uniform mapping between these perceptions and specific trust dimensions. Anthropomorphic design may make trust more resistant to reductions in system reliability [6], while agent appearance and reliability can jointly shape trustworthiness perceptions [20]. Other social interface cues have produced effects concentrated in benevolence, without corresponding changes in ability or integrity [22]. These findings justify separating cognitive and social perceptions and retaining multiple trust dimensions, but they do not support assuming that every perception will relate equally to ability, integrity, and benevolence.

### 2.4 AI Assistants Embedded in Platform Contexts

AI assistants increasingly appear within platforms where users are already reading and responding to content. Research on AI-mediated communication examines how computational systems intervene in communication between people [14]. Believing that AI contributed to a message can alter judgments of its author's trustworthiness [19]; agents participating in online communities can acquire meaning through local norms and repeated participation [44]; and AI-generated summaries embedded in simulated social-media discussions can influence opinion formation [12].

These studies suggest that the consequences of AI assistance depend on the surrounding communication environment. However, they primarily examine AI involvement in interpersonal messages, repeated community participation, or downstream opinion formation. Less is known about the immediate judgments users form about an embedded assistant itself during a single platform encounter. The present study focuses on this immediate evaluative context; details of the controlled platform implementation are provided in Section 4.

## 3 Hypothesis Development and Research Question

The proposed relationships follow a progression from experimentally assigned interface cues to capability perceptions, anthropomorphism, and differentiated trust. We use directional hypotheses where prior theory and evidence support a specific prediction. We use a research question where the literature supports examining a relationship but does not support predicting a particular pattern. Additional data-driven and supplementary analyses are identified as exploratory in the Data Analysis section.

### 3.1 Multifunctionality as a Capability Cue

Under the cue-route perspective reviewed in Section 2.1, visible interface features provide evidence about a system's capabilities, but users may interpret the same evidence differently. We therefore distinguish between the functionality displayed by the interface (assigned multifunctionality) and the functional breadth attributed to the assistant by users (perceived multifunctionality). Whether displayed functional breadth translates into perceived functional breadth is itself an important question in this study.

Participants assigned to the multifunctional condition encountered three distinct functions, whereas those assigned to the single-function condition encountered only one. Although individuals may differ in how fully they take up this difference, the multifunctional display provides more visible evidence of functional breadth and should raise perceived multifunctionality on average. Perceived functional breadth may, in turn, be associated with broader judgments of cognitive capability. When users perceive that an assistant can perform a wider range of functions, they may regard it as possessing greater knowledge, flexibility, or problem-solving capacity. Research spanning physical robots, software robots, and chatbots identifies multifunctionality as an antecedent of perceived intelligence [31], while survey research links perceived multifunctionality with perceived product smartness [25]. Accordingly, we predict:

- **H1:** Participants assigned to the multifunctional condition will report higher perceived multifunctionality than participants assigned to the single-function condition.
- **H2:** Perceived multifunctionality will be positively associated with perceived intelligence.

### 3.2 Explanation as a Capability Cue

Whereas multifunctionality displays the range of functions an assistant can perform, an explanation can provide information about the apparent reasoning behind its outputs. Under the cue-route perspective, making this reasoning basis visible may give users additional evidence from which to judge the assistant's cognitive capability. If users interpret the rationale as indicating that the assistant can connect its outputs to reasons, they may evaluate the assistant as more knowledgeable, thoughtful, or intelligent [13, 50].

Prior evidence makes this directional prediction plausible. Greater agent transparency has increased perceived intelligence [53], and perceived explainability has been positively associated with cognitive evaluations of automated systems [54]. Although the value of a particular explanation may depend on its content and audience, providing a rationale makes an additional potential cue of cognitive capability available. We therefore predict:

**H3:** Participants assigned to the explanation-present condition will report higher perceived intelligence than participants assigned to the explanation-absent condition.

### 3.3 Perceived Intelligence and Anthropomorphism

Perceived intelligence and anthropomorphism represent different evaluations of an AI assistant. The former concerns cognitive competence, whereas the latter concerns personlike characteristics, motivations, or intentions. Theoretical and empirical work reviewed in Section 2.2 suggests that observable evidence of cognitive capability can contribute to humanlike interpretations of an agent. Perceived intelligence and anthropomorphism were strongly and positively associated in prior research on a virtual teaching assistant [49] and positively related in earlier work on AI assistants [33]. We therefore predict:

**H4:** Perceived intelligence will be positively associated with anthropomorphism.

### 3.4 Perceived Intelligence, Anthropomorphism, and Differentiated Trust

As reviewed in Section 2.3, perceived intelligence and anthropomorphism may correspond to different bases of trust, but existing findings do not establish a consistent dimension-specific profile. Perceived intelligence may be especially relevant to judgments of the assistant's competence, whereas anthropomorphism may be more relevant to relational judgments. However, the literature does not justify predicting identical or fully distinct relationships across ability, integrity, and benevolence. We therefore retain the three dimensions as separate latent outcomes and ask:

**RQ1:** When perceived intelligence and anthropomorphism are estimated together, how are they differentially associated with ability-, integrity-, and benevolence-based trust?

Figure 1 summarizes the hypothesized relationships in H1–H4 and the differentiated trust relationships examined in RQ1.

### 3.5 Supplementary Exploratory Analysis of Randomized Effects

H1 and H3 specify the randomized effects most directly implied by the two interface manipulations. To provide a complete account of the factorial experiment, we additionally estimate the main effects of assigned multifunctionality and assigned explanation, together with their interaction, across all six latent outcomes. Effects corresponding to H1 and H3 are interpreted in relation to the hypotheses above. All remaining randomized paths and multifunctionality-by-explanation interactions are treated as supplementary exploratory analyses rather than as an additional numbered research question.

## 4 Methodology

### 4.1 Experiment Design

This study used a randomized $2 \times 2$ between-participants design with two experimentally assigned factors: AI multifunctionality (single-function vs. multifunctional) and explanation (explanation-present vs. explanation-absent).

- **AI multifunctionality.** This factor manipulated the number of functions that the interface presented the AI assistant as performing. In the *single-function* condition, the interface displayed one function by summarizing the

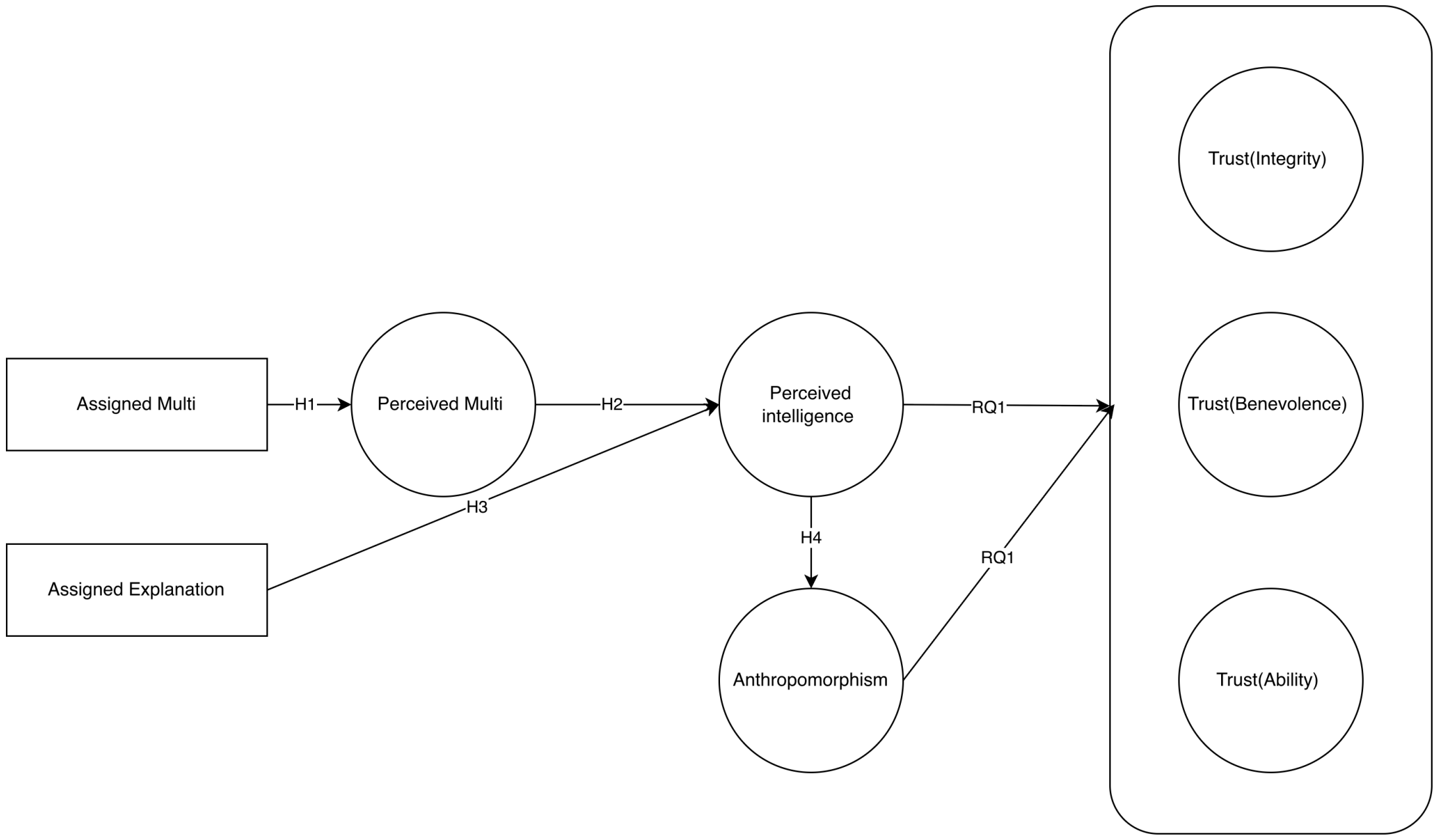

Fig. 1. **Hypothesized research model.** Solid paths represent the focal relationships examined in the study. Paths labeled H1–H4 correspond to the four hypotheses, whereas the paths from perceived intelligence and anthropomorphism to ability-, integrity-, and benevolence-based trust correspond to RQ1. Rectangles represent experimentally assigned variables, and ellipses represent latent constructs.

Reddit post. In the *multifunctional* condition, the interface displayed three functions: topic tagging, summarization, and fact-checking. The assistant header also explicitly identified the display as containing one or three functions.

- **Explanation.** This factor manipulated whether the interface provided a rationale for the pre-generated AI outputs. In the *explanation-present* condition, a "Why this analysis" panel briefly explained why the assistant had produced its outputs. In the *explanation-absent* condition, the assistant presented the outputs alone.

Crossing these factors yielded four experimental conditions. The single-function and multifunctional assistants each appeared with either the explanation-present or explanation-absent interface.

The experiment was implemented using *Outer Limits*, a Chrome extension that presents AI assistance while users browse Reddit [32]. Before data collection, Anthropic's Claude 4.8 was used to generate the HTML/CSS implementation of the assistant panel, following common visual conventions of embedded AI-assistant interfaces. Claude 4.8 was also used to generate draft assistant outputs: topic tags, a summary, a fact-check, and an explanatory rationale. The research team reviewed and finalized both the interface implementation and the generated text, including checking the outputs for factual accuracy. The resulting panel was a custom study interface rather than the native interface of Claude or another commercial AI product.

No live model generation occurred during study sessions. Instead, the extension rendered the corresponding pre-generated outputs locally according to each participant's randomized condition. This procedure ensured that participants assigned to the same condition encountered identical content. Visually, the panel was labeled "AI Assistant" and indicated

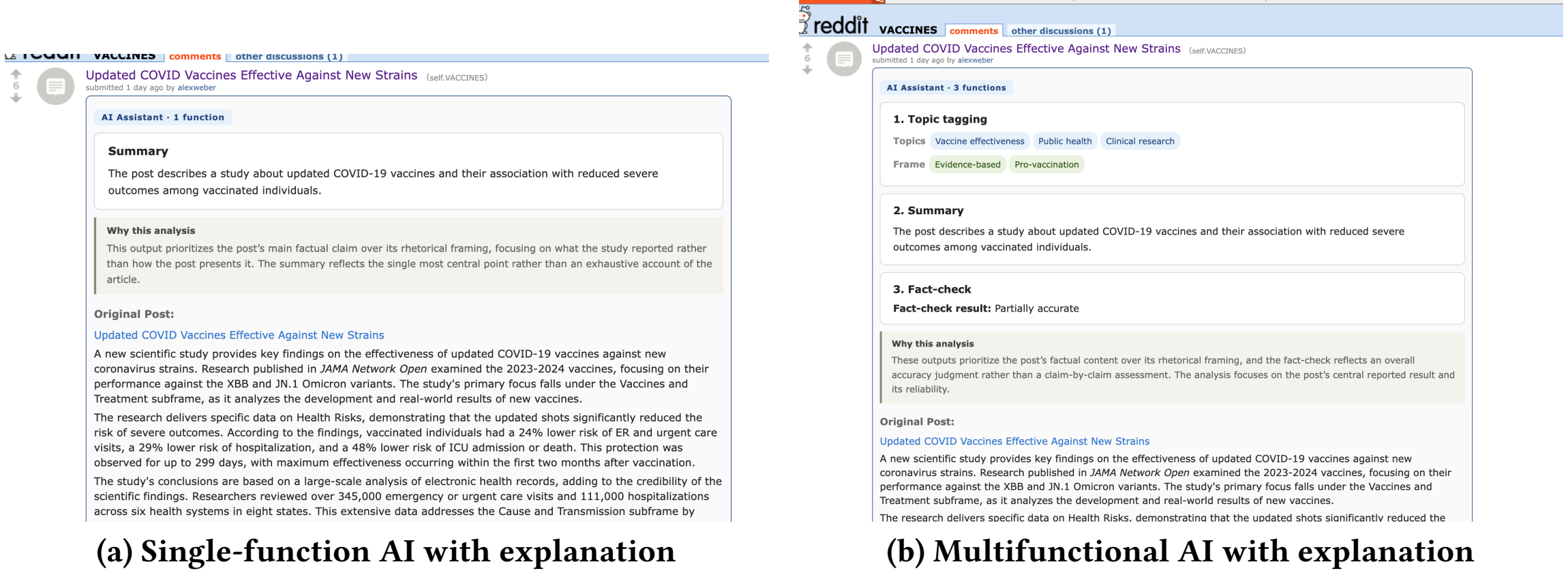

**(a) Single-function AI with explanation** **(b) Multifunctional AI with explanation**

Fig. 2. **Examples of the Outer Limits interface under the two assigned-multifunctionality conditions.** The single-function assistant provided one function (a summary), whereas the multifunctional assistant provided three functions (topic tagging, a summary, and a fact-check). Both panels show the explanation condition, indicated by the "Why this analysis" panel, and the same Reddit post was displayed across conditions.

whether it contained one or three functions. Each function appeared as a distinct output section; when assigned, the explanatory rationale appeared in a separate "Why this analysis" panel.

### 4.2 Experiment Procedure

Participants were told that they would be testing *Outer Limits*, a new Chrome-based AI assistant for Reddit, and that the study would evaluate whether the extension functioned as intended and whether users found it useful. They first completed a pre-survey containing informed consent, eligibility screening, their Prolific ID, and demographic questions. Eligibility required participants to be at least 18 years old, reside in the United States, use the Chrome browser, have an existing Reddit account, and be logged into Reddit before beginning the study. Reddit authentication occurred directly through Reddit; the study did not collect or process participants' login credentials. The logged-in Old Reddit page supplied the interface host required by *Outer Limits*; participants then installed the extension and were randomly assigned to one of the four experimental conditions.

The stimulus concerned a Reddit discussion of an article about updated COVID-19 vaccines and severe health outcomes. The extension opened the same designated Reddit URL for all participants in the Old Reddit interface, which was required for compatibility with *Outer Limits*. The extension then locally rendered a preconstructed version of the page corresponding to each participant's assigned condition.

The research team prepared the simulated Reddit discussion before data collection. The displayed post content was a ChatGPT-generated summary of the source news article about updated COVID-19 vaccines. The three displayed comments and the fictional usernames for the post author and commenters were also generated with ChatGPT rather than taken from real Reddit users. The research team reviewed and finalized the post, comments, and usernames before the experiment. Across all four conditions, the displayed post title, author username, posting time, post content, post score, commenter usernames, comments, comment timestamps, and comment scores were identical. The only differences were the two experimentally assigned features of the AI assistant: whether it performed one or three functions and whether it displayed a rationale for its outputs. The prepared content was rendered locally and was not published to Reddit, keeping the simulated discussion separate from the live Reddit community.

Figure 2 shows examples of the experimental interface rendered from the same underlying Reddit URL.

After reviewing the designated Reddit post, participants followed the study link to a post-survey measuring perceived multifunctionality, perceived intelligence, anthropomorphism, trust, AI literacy, and manipulation checks. Pre-survey, system, and post-survey records were linked using participants' Prolific IDs.

### 4.3 Ethical Considerations

Participants provided informed consent before beginning the study. To preserve experimental realism, the protocol temporarily concealed the study's specific hypotheses and the simulated nature of the displayed Reddit content. Participants were told that they were evaluating the AI-assistant features of *Outer Limits* for Reddit. They were led to believe that the post, comments, engagement indicators, and AI outputs formed part of a live Reddit encounter and that the assistant had produced its outputs during that encounter. In fact, the simulated post summary, comments, and fictional usernames had been generated with ChatGPT, whereas the AI-assistant panel's HTML/CSS and its output content had been generated with Claude 4.8. The research team reviewed and finalized all generated materials before data collection. The extension rendered the corresponding pre-generated content locally rather than calling a model during the study. This arrangement held the background Reddit content constant across conditions, standardized the AI outputs within each condition, and prevented study activity from affecting the Reddit community. Comments and other actions entered by participants in the simulated interface were recorded only by the study system and were not submitted to Reddit. Participants authenticated directly through Reddit, and the study did not collect or process their Reddit login credentials. At the end of the session, participants were fully debriefed about the simulated content, the experimental manipulations, and the reason for the temporary concealment. Participants could discontinue their participation at any time. The Institutional Review Board at Lehigh University approved all study procedures.

### 4.4 Recruitment and Participants

Participants were recruited through Prolific, and data were collected between July 13 and July 22, 2026. The study was advertised as taking approximately 30 minutes, with compensation of \$6.50. The sample-size calculation targeted the randomized $2 \times 2$ design. For a fixed-effects factorial ANOVA, a one-degree-of-freedom interaction with Cohen's $f = .25$, $\alpha = .05$, and 95% power required 210 participants. Data were obtained from 283 participants. Eleven participants were excluded for failing the study's attention-check criterion, and two were excluded because system records indicated that *Outer Limits* had not successfully rendered the assigned manipulated post. The final analytic sample therefore consisted of 270 participants. The retained cell sizes were 67 in the single-function/no-explanation condition, 64 in the single-function/explanation condition, 65 in the multifunctional/no-explanation condition, and 74 in the multifunctional/explanation condition. The retained sample exceeded the design-stage target; the latent structural estimates are reported with confidence intervals and specification checks.

The final sample comprised 140 women (51.9%), 128 men (47.4%), and two non-binary participants (0.7%). Ages ranged from 18 to 67 years ($M = 35.71$, $SD = 11.17$). The largest education groups were bachelor's degree (35.2%), some college without a degree (22.2%), high-school graduate (15.6%), associate degree (10.7%), and master's degree (9.6%). The mutually exclusive race/ethnicity categories used for analysis were White/European American (55.2%), Black/African American (18.5%), multiracial (14.4%), Asian/Asian American (6.3%), Hispanic/Latinx (4.4%), and other single-race identities (1.1%). Objective AI-literacy scores ranged from 6 to 17 out of a possible 17 ($M$ = 13.46, $SD$ = 2.06, median = 14).

No statistically significant differences were detected across the four experimental conditions in age, gender, education, race/ethnicity, household income, or objective AI literacy (all $p > .05$).

### 4.5 Experimental Manipulations

*4.5.1 Assigned Multifunctionality.* Assigned multifunctionality had two levels. In the single-function condition, the interface displayed one distinct function on the Reddit post. In the multifunctional condition, the interface displayed multiple distinct functions. The post-survey included both a perceived-multifunctionality scale and a factual manipulation-check item asking how many distinct functions the assistant had performed.

To assess whether participants recognized the assigned multifunctionality manipulation, the post-survey asked, How many distinct functions did the AI assistant perform on this post? Participants selected either "One function" or 'More than one function'. Recognition differed substantially by assigned condition. Among participants assigned to the multifunctional condition, 134 of 139 (96.4%) selected "More than one function". Among those assigned to the single-function condition, 109 of 131 (83.2%) selected "One function". The manipulation check indicated that the multifunctionality manipulation was successfully recognized: recognition responses were strongly associated with assigned condition, $\chi^2(1) = 175.20$, $p < .001$, Cramér's $V = .806$. Overall, 243 of 270 participants (90.0%) correctly identified the displayed functional breadth. Participants were analyzed according to their randomized condition regardless of their response to this post-treatment recognition item.

*4.5.2 Assigned Explanation.* Assigned explanation also had two levels. In the explanation-present condition, the AI presented a rationale for why it had produced its outputs, beyond presenting the outputs themselves. In the explanation-absent condition, the AI presented the outputs alone. To assess recognition, the post-survey asked, Did the AI give any explanation of why it produced those outputs (beyond the outputs themselves)? Participants selected either "Yes" or "No". Among participants assigned to the explanation-present condition, 104 of 138 (75.4%) selected "Yes". Among those assigned to the explanation-absent condition, 119 of 132 (90.2%) selected "No". The manipulation check indicated that the explanation manipulation was successfully recognized overall, although recognition was imperfect: recognition responses were strongly associated with assigned condition, $\chi^2(1) = 117.92$, $p < .001$, Cramér's $V = .661$. Overall, 223 of 270 participants (82.6%) correctly identified whether a rationale had been displayed. Participants were analyzed according to their randomized condition regardless of their response to this post-treatment recognition item.

### 4.6 Variables and Measurements

The full wording, response options, and scale anchors for all measures are provided in Appendix A.

*4.6.1 Perceived Multifunctionality.* To measure perceived multifunctionality, we adapted a four-item scale from Lee and Shin [25] and modified the wording to refer to the AI assistant used in this study. Example items include "This AI assistant has multiple functions" and "This AI assistant performs multiple tasks." All items were measured on a seven-point scale ranging from 1 (*strongly disagree*) to 7 (*strongly agree*). The scale demonstrated high internal consistency ($\alpha$ = .957, $M$ = 4.60, $SD$ = 1.66). Perceived multifunctionality captured participants' post-treatment perception of the assistant, so the four items formed a latent construct and randomized assignment entered the model separately.

4.6.2 *Perceived Intelligence.* Perceived intelligence was measured using the five-item perceived-intelligence dimension of the Godspeed questionnaire [2], with the referent adapted to the AI assistant. The original semantic-differential pairs were retained (e.g., "Incompetent/Competent" and "Unintelligent/Intelligent"). Items were measured on five-point scales ranging from the adjective shown on the left (1) to the adjective shown on the right (5). The scale demonstrated high internal consistency ($\alpha$ = .936, $M$ = 3.84, $SD$ = .89), and the five items were modeled as indicators of a latent perceived-intelligence factor.

4.6.3 *Anthropomorphism.* To measure anthropomorphism, we adapted the four-item measure from Chen et al. [4], also adapted to service agents by Choi et al. [5]. We changed the referent to the AI assistant and made minor wording adjustments. Example items include "The AI assistant feels like a person" and "The AI assistant has its own personality." We used a seven-point response scale rather than the original nine-point scale, with the same verbal anchors, *strongly disagree* (1) and *strongly agree* (7). The scale demonstrated high internal consistency ($\alpha$ = .918, $M$ = 2.61, $SD$ = 1.53), and the four items were modeled as indicators of a latent anthropomorphism factor.

4.6.4 *Trust.* Trust was measured using items adapted from the ability, benevolence, and integrity framework used by Wang and Benbasat [50]. The wording was modified to refer to the AI assistant and its assistance with the Reddit post. Benevolence was measured with three items (e.g., "This AI assistant has my best interests in mind"; $\alpha$ = .913, $M$ = 4.14, $SD$ = 1.54). Integrity was measured with three items (e.g., "This AI assistant is honest"; $\alpha$ = .876, $M$ = 4.81, $SD$ = 1.31). Ability was measured with five items (e.g., "This AI assistant has the expertise needed to analyze the Reddit post"; $\alpha$ = .950, $M$ = 5.40, $SD$ = 1.21). All items used a seven-point scale ranging from 1 (*strongly disagree*) to 7 (*strongly agree*). Because the study examines conceptually distinct forms of trust, benevolence, integrity, and ability were modeled as separate latent factors. Residual covariances among the three factors were freely estimated.

4.6.5 *Objective AI Literacy.* Objective AI literacy was measured using a 17-item multiple-choice knowledge test adapted from Tully et al. [48]. The items assessed knowledge of AI applications, machine learning, data interpretation, sensing, programmability, and ethical issues (e.g., "Which of the following is NOT powered by AI?"). Each correct response received one point, and responses were summed to produce a score ranging from 0 to 17, with higher scores indicating greater objective AI knowledge. Observed scores ranged from 6 to 17 ($M$ = 13.46, $SD$ = 2.06, median = 14).

### 4.7 Data Analysis Methods

All analyses were conducted in R using `lavaan` [43]. We first estimated a six-factor measurement model comprising perceived multifunctionality, perceived intelligence, anthropomorphism, benevolence, integrity, and ability.

We then estimated a latent-outcome factorial model to examine the effects of the two randomized cues across the six focal outcomes. Each outcome was regressed on assigned multifunctionality, assigned explanation, and their interaction. We report raw $p$ values and Benjamini–Hochberg-adjusted [3] $q$ values across the 12 randomized main-effect tests. The six factorial-interaction tests were adjusted as a separate family.

The structural SEM reported in the Findings contained all paths in the hypothesized model shown in Figure 1, together with an exploratory objective-AI-literacy extension. A screen of literacy moderation across the 10 structural paths represented in the proposed model identified only the interaction between assigned multifunctionality and objective AI literacy after Benjamini–Hochberg correction. This interaction and the objective-literacy main effect were therefore included in the reported SEM.

The SEM was estimated using MLR with FIML, standardized latent-variable scaling, and fixed randomized predictors. For RQ1, omnibus Wald tests compared the three standardized trust-path coefficients within each predictor using their robust delta-method covariance matrix. Pairwise comparisons followed significant omnibus tests, with Benjamini–Hochberg correction applied separately within each set of three comparisons. As a robustness check, we re-estimated the same model using WLSMV [42], treating the 24 indicators as ordered categorical variables. H1–H4 and RQ1 were focal; the literacy moderation and WLSMV analysis were exploratory and robustness analyses, respectively. Randomized effects permit causal interpretation, whereas paths among concurrently measured post-exposure judgments remain associational.

## 5 Findings

All 270 retained participants had complete data on the focal measurement indicators, objective AI literacy score, and experimental-condition variables. We first report the measurement model, followed by the total randomized effects of the interface manipulations, the structural model, the objective-AI-literacy moderation, and robustness analyses.

### 5.1 Measurement Model and Discriminant Validity

The six-factor confirmatory factor analysis included perceived multifunctionality, perceived intelligence, anthropomorphism, benevolence, integrity, and ability. It showed good fit, robust $\chi^2(237) = 425.82$, $p < .001$, CFI = .968, TLI = .963, RMSEA = .056, 90% CI [.047, .065], and SRMR = .043. Standardized loadings ranged from .738 to .943, composite reliabilities from .885 to .958, and AVE from .721 to .851. Each of the 15 construct pairs passed the Fornell–Larcker [11] criterion and remained below the .90 HTMT threshold [15]; the largest HTMT value was .676 for ability and integrity. Appendix B reports the pairwise latent correlations and the Fornell–Larcker and HTMT discriminant-validity results.

### 5.2 Experimental Effects on Latent Outcomes

We tested whether assigned multifunctionality and assigned explanation affected each of the six latent outcomes. We also tested their interaction—that is, whether the effect of multifunctionality differed between the explanation and no-explanation conditions. The latent factorial model showed good fit, robust $\chi^2(291) = 497.85$, $p < .001$, CFI = .966, TLI = .959, RMSEA = .052, 90% CI [.044, .061], and SRMR = .040.

As shown in Table 1, assigned multifunctionality significantly increased perceived multifunctionality ($\beta = .402$, $p < .001$, $q < .001$). The effect of assigned multifunctionality on perceived intelligence was small and did not survive correction for multiple testing ($\beta = .128$, $p = .043$, $q = .172$). Assigned multifunctionality did not significantly affect anthropomorphism or any of the three trust dimensions.

Assigned explanation did not significantly affect perceived intelligence ($\beta = -.018$, $p = .770$). Although its estimate for benevolence was positive ($\beta = .147$, $p = .021$), this effect did not remain significant after correction for multiple testing ($q = .123$). Its effects on perceived multifunctionality, anthropomorphism, integrity, and ability were also nonsignificant.

None of the multifunctionality-by-explanation interactions was significant ($|\beta| = .003$–$.038$, all $p > .53$). We therefore found no evidence that the effect of assigned multifunctionality differed depending on whether the assistant provided an explanation. Overall, displaying multiple functions increased perceived multifunctionality, and this was the only randomized main effect that survived correction for multiple testing.

| Latent outcome | Multifunctionality $\beta$ | Multifunctionality $p$ | Multifunctionality $q_{\text{main}}$ | Explanation $\beta$ | Explanation $p$ | Explanation $q_{\text{main}}$ | Interaction $\beta$ | Interaction $p$ | Interaction $q_{\text{int}}$ |
|---|---|---|---|---|---|---|---|---|---|
| Perceived multifunctionality | .402 | < .001 | < .001 | .018 | .761 | .840 | −.025 | .665 | .967 |
| Perceived intelligence | .128 | .043 | .172 | −.018 | .770 | .840 | −.003 | .967 | .967 |
| Anthropomorphism | .007 | .913 | .913 | .089 | .157 | .270 | .025 | .686 | .967 |
| Benevolence | .040 | .530 | .793 | .147 | .021 | .123 | −.035 | .576 | .967 |
| Integrity | .090 | .157 | .270 | .103 | .106 | .255 | −.009 | .890 | .967 |
| Ability | .111 | .072 | .216 | .033 | .595 | .793 | −.038 | .536 | .967 |

Table 1. **Experimental effects on the six latent outcomes.** Each outcome was simultaneously regressed on effect-coded assigned multifunctionality, effect-coded assigned explanation, and their interaction. The interaction tests whether the effect of multifunctionality differed between the explanation and no-explanation conditions. Benjamini–Hochberg $q_{\text{main}}$ values adjust across the 12 main-effect tests, whereas $q_{\text{int}}$ values adjust across the six interaction tests.

## 5.3 Structural Associations among Capability, Anthropomorphism, and Trust

We estimated the hypothesized structural model, augmented with the exploratory objective-AI-literacy terms, using robust maximum likelihood (MLR). The model showed acceptable overall fit, robust $\chi^2(333) = 593.22$, $p < .001$, CFI = .958, TLI = .953, RMSEA = .054, 90% CI [.047, .062], SRMR = .063, AIC = 16,924.20, and BIC = 17,237.27. Figure 3 presents the model, and Table 2 reports the estimated structural paths.

The results were consistent with H1 and H2. Participants assigned to the multifunctional condition reported higher perceived multifunctionality than participants assigned to the single-function condition ($\beta = .392$, $p < .001$), consistent with H1. Perceived multifunctionality was, in turn, positively associated with perceived intelligence ($\beta = .595$, $p < .001$), consistent with H2. H3 was not supported: providing an explanatory rationale did not increase perceived intelligence ($\beta = -.025$, $p = .634$).

Perceived intelligence was positively associated with anthropomorphism ($\beta = .393$, $p < .001$), consistent with H4. Perceived multifunctionality, perceived intelligence, and anthropomorphism were measured in the same post-exposure survey.

RQ1 asked whether perceived intelligence and anthropomorphism had different associations with ability-, integrity-, and benevolence-based trust. Perceived intelligence was positively associated with benevolence ($\beta = .364$, $p < .001$), integrity ($\beta = .493$, $p < .001$), and ability ($\beta = .651$, $p < .001$). Anthropomorphism had an additional positive association with benevolence ($\beta = .335$, $p < .001$), but not with integrity ($\beta = .071$, $p = .262$) or ability ($\beta = .015$, $p = .758$).

The standardized anthropomorphism coefficients differed across the three trust dimensions, Wald $\chi^2(2) = 28.00$, $p < .001$. The association with benevolence was larger than those with integrity and ability (both BH-adjusted $q < .001$), whereas the integrity–ability comparison was not significant ($q = .379$). The standardized perceived-intelligence coefficients also differed jointly, Wald $\chi^2(2) = 18.28$, $p < .001$. The association with ability was larger than those with benevolence ($q < .001$) and integrity ($q = .034$), and the association with integrity was larger than that with benevolence ($q = .034$).

These results provide a differentiated answer to RQ1. Perceived intelligence was associated with all three trust dimensions, whereas the incremental association of anthropomorphism was specific to benevolence. The perceived-intelligence-to-ability path was the largest of the trust paths. This result should nevertheless be interpreted cautiously because perceived intelligence and ability-based trust are conceptually related, although the measurement analysis supported their empirical distinctiveness.

The structural model also included an exploratory interaction between assigned multifunctionality and objective AI literacy. This interaction was positive and significant ($b = .604$, 95% CI [.311, .897], $\beta = .264$, $p < .001$), indicating that the effect of assigned multifunctionality on perceived multifunctionality increased with objective AI literacy. The moderation-screening procedure and conditional effects are reported in Section 5.4.

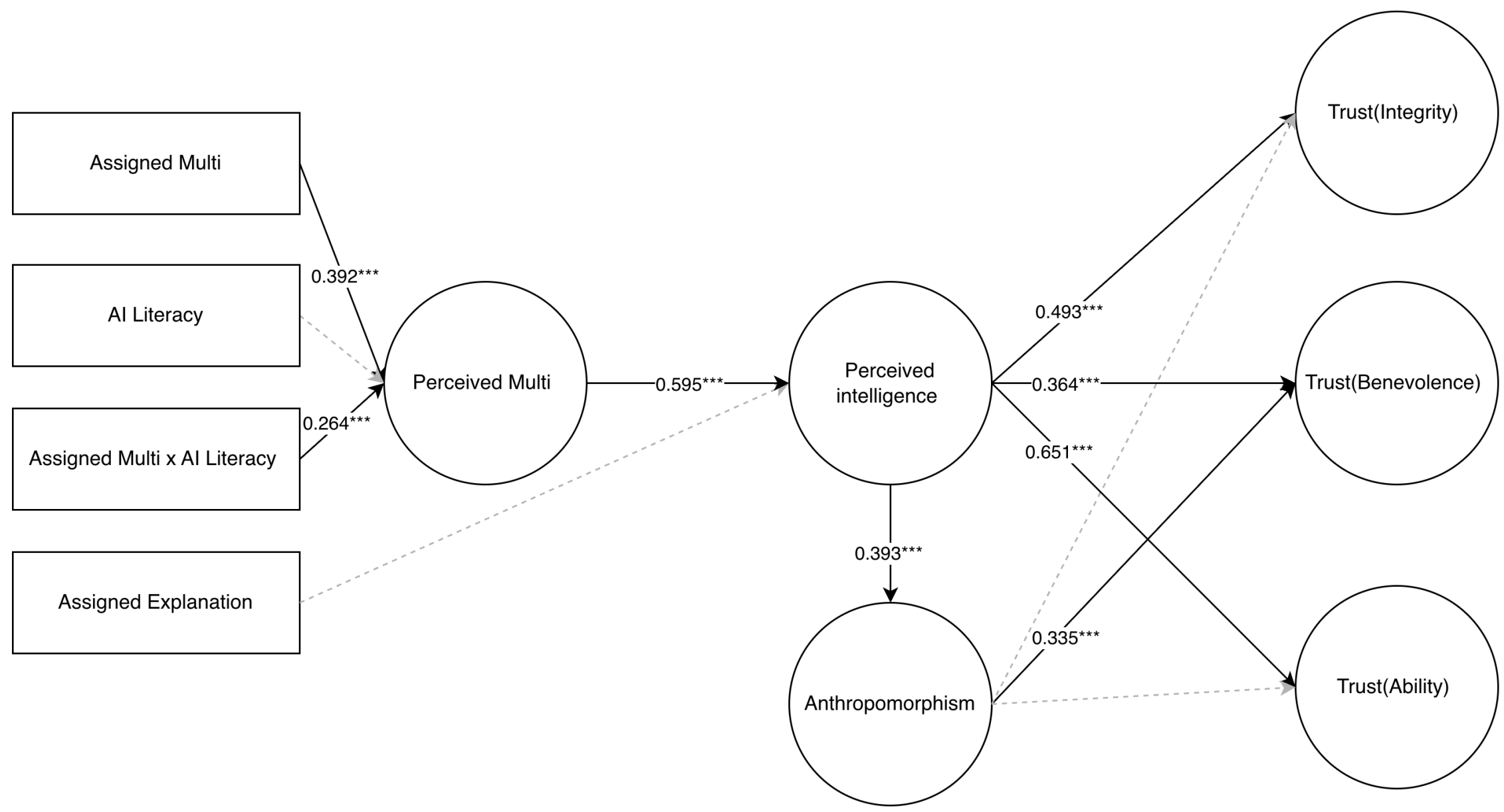

Fig. 3. **Results of the hypothesized structural model augmented with the exploratory objective-AI-literacy terms and estimated using MLR.** Rectangles represent observed variables and ellipses represent latent constructs. Solid black arrows indicate statistically significant paths, whereas dashed gray arrows indicate nonsignificant paths. Values are standardized coefficients ($\beta$). Residual covariances among the three trust dimensions were estimated but are omitted from the diagram. Paths among post-exposure perceptions represent concurrent structural associations and do not establish temporal or causal order. $^{***}p < .001$, $^{**}p < .01$, $^{*}p < .05$.

| **Path** | $b$ | $SE$ | $\beta$ | **95% CI for** $b$ | $p$ |
|---|---|---|---|---|---|
| Assigned multifunctionality → Perceived multifunctionality | .895 | .138 | .392 | [.624, 1.165] | $< .001^{***}$ |
| Objective AI literacy → Perceived multifunctionality | −.088 | .073 | −.077 | [−.230, .055] | .229 |
| Multifunctionality × objective literacy → Perceived multifunctionality | .604 | .150 | .264 | [.311, .897] | $< .001^{***}$ |
| Perceived multifunctionality → Perceived intelligence | .650 | .103 | .595 | [.448, .851] | $< .001^{***}$ |
| Assigned explanation → Perceived intelligence | −.061 | .129 | −.025 | [−.314, .191] | .634 |
| Perceived intelligence → Anthropomorphism | .344 | .047 | .393 | [.251, .436] | $< .001^{***}$ |
| Perceived intelligence → Benevolence | .361 | .070 | .364 | [.224, .497] | $< .001^{***}$ |
| Anthropomorphism → Benevolence | .380 | .072 | .335 | [.240, .520] | $< .001^{***}$ |
| Perceived intelligence → Integrity | .465 | .080 | .493 | [.309, .621] | $< .001^{***}$ |
| Anthropomorphism → Integrity | .077 | .069 | .071 | [−.058, .212] | .262 |
| Perceived intelligence → Ability | .693 | .097 | .651 | [.503, .883] | $< .001^{***}$ |
| Anthropomorphism → Ability | .019 | .060 | .015 | [−.099, .136] | .758 |

Table 2. Structural paths in the hypothesized MLR model, augmented with the exploratory objective-AI-literacy main effect and interaction. Unstandardized coefficients are denoted by $b$ and standardized coefficients by $\beta$. $^{***}p < .001$, $^{**}p < .01$, $^{*}p < .05$.

### 5.4 Exploratory Moderation by Objective AI Literacy

We examined whether objective AI literacy moderated each of the 10 structural relationships represented in the hypothesized model. Only the interaction between assigned multifunctionality and objective AI literacy in predicting perceived multifunctionality remained significant after Benjamini–Hochberg correction across these 10 tests ($q < .001$). We therefore incorporated this interaction, together with its component main effects, into the latent structural model.

Objective AI literacy positively moderated the effect of assigned multifunctionality on perceived multifunctionality, $b = .604$, 95% CI [.311, .897], $\beta = .264$, $p < .001$. Specifically, the difference in perceived multifunctionality between the assigned multifunctional and single-function conditions became larger as participants' objective AI literacy increased. Because the location of the moderation was identified and estimated using the same sample, this result should be treated as exploratory and confirmed in an independent sample.

### 5.5 Ordinal-Indicator Robustness Check

Because the focal constructs were measured using ordered rating-scale indicators, including the five-category Godspeed items, we examined whether treating these indicators as approximately continuous in the primary MLR model affected the results. We therefore re-estimated the same measurement and structural specification using weighted least squares mean- and variance-adjusted estimation (WLSMV), treating all 24 indicators as ordered categorical variables. The ordinal model showed acceptable fit, scaled $\chi^2(333) = 755.34$, $p < .001$, CFI = .980, TLI = .983, RMSEA = .069, 90% CI [.062, .075], and SRMR = .059. Table 3 reports the structural-path estimates from the WLSMV model.

The WLSMV analysis reproduced the conclusions corresponding to H1–H4. Assigned multifunctionality remained positively associated with perceived multifunctionality ($\beta = .260$, $p < .001$), consistent with H1. Perceived multifunctionality remained positively associated with perceived intelligence ($\beta = .719$, $p < .001$), consistent with H2. The association between assigned explanation and perceived intelligence remained nonsignificant ($\beta = .093$, $p = .150$), again providing no support for H3. Perceived intelligence remained positively associated with anthropomorphism ($\beta = .547$, $p < .001$), consistent with H4. The exploratory assigned-multifunctionality-by-objective-literacy interaction also remained significant ($\beta = .268$, $p < .001$).

The WLSMV results also produced the same substantive answer to RQ1. Perceived intelligence was positively associated with benevolence ($\beta = .467$, $p < .001$), integrity ($\beta = .605$, $p < .001$), and ability ($\beta = .776$, $p < .001$). Anthropomorphism retained an additional positive association with benevolence ($\beta = .297$, $p < .001$), but not with integrity ($\beta = -.053$, $p = .437$) or ability ($\beta = -.126$, $p = .055$). Thus, the differentiated trust profile and the perceived-intelligence-to-anthropomorphism association were stable across the MLR and WLSMV estimators.

*Interpretive scope.* Because the experimental conditions were randomly assigned, the effects of assigned multifunctionality and assigned explanation can be interpreted causally within this experiment. In contrast, the post-exposure latent constructs were measured concurrently; paths among them therefore represent associations consistent with, but not evidence of, the proposed causal ordering.

## 6 Discussion

How do users evaluate the capabilities of a platform-embedded AI assistant, and how do these evaluations relate to different dimensions of trust? In our study, assignment to the multifunctional condition increased perceived functional breadth relative to the single-function condition. In participants' post-exposure evaluations, perceived intelligence was associated with all three trust dimensions. After accounting for perceived intelligence, the additional association of anthropomorphism was concentrated in benevolence, or whether the assistant appeared to have the user's interests in mind. We begin with how users interpret the displayed functions, then consider how capability judgments and personlike attribution relate to different bases of trust.

| **Path** | $b$ | $SE$ | $\beta$ | **95% CI for** $b$ | $p$ |
|---|---|---|---|---|---|
| Assigned multifunctionality → Perceived multifunctionality | .559 | .157 | .260 | [.251, .868] | $< .001^{***}$ |
| Objective AI literacy → Perceived multifunctionality | −.023 | .077 | −.021 | [−.173, .128] | .768 |
| Multifunctionality × objective literacy → Perceived multifunctionality | .579 | .156 | .268 | [.272, .886] | $< .001^{***}$ |
| Perceived multifunctionality → Perceived intelligence | .966 | .082 | .719 | [.805, 1.127] | $< .001^{***}$ |
| Assigned explanation → Perceived intelligence | .269 | .187 | .093 | [−.097, .636] | .150 |
| Perceived intelligence → Anthropomorphism | .451 | .051 | .547 | [.351, .551] | $< .001^{***}$ |
| Perceived intelligence → Benevolence | .438 | .064 | .467 | [.312, .563] | $< .001^{***}$ |
| Anthropomorphism → Benevolence | .337 | .072 | .297 | [.195, .479] | $< .001^{***}$ |
| Perceived intelligence → Integrity | .512 | .063 | .605 | [.388, .636] | $< .001^{***}$ |
| Anthropomorphism → Integrity | −.054 | .070 | −.053 | [−.191, .083] | .437 |
| Perceived intelligence → Ability | .766 | .074 | .776 | [.620, .911] | $< .001^{***}$ |
| Anthropomorphism → Ability | −.151 | .079 | −.126 | [−.305, .003] | .055 |

Table 3. WLSMV robustness analysis of the hypothesized structural paths and exploratory objective-AI-literacy terms included in the primary MLR model.

### 6.1 Recognizing functional breadth and judging intelligence

Displaying three functions rather than one increased perceived multifunctionality ($\beta = .402$, $q < .001$), with 90.0% of participants correctly recognizing their assigned functional breadth. The corresponding effect on perceived intelligence was smaller ($\beta = .128$, $p = .043$, $q = .172$); perceived multifunctionality was the only outcome with a randomized main effect that survived correction. Within the structural model, however, perceived multifunctionality was strongly associated with perceived intelligence ($\beta = .595$, $p < .001$).

This contrast matters for how capability displays are evaluated. Sundar's cue-route account describes how users draw on visible interface features when forming judgments [46]. Here, the same experiment that established uptake of a functional-breadth cue supplied a more qualified answer about its value as evidence of intelligence. A participant who regards an assistant as multifunctional may also regard it as intelligent, while adding functions to its display produces a more modest change in that broader judgment. The two estimates concern different relationships and can coexist. Displaying more functions and making an assistant seem more intelligent are separate design achievements. Measuring both in the same experiment reveals which judgment a capability display actually changes, even when those judgments are closely associated in users' evaluations.

The exploratory literacy result makes this distinction more specific. The effect of assigned multifunctionality on perceived multifunctionality increased with objective AI literacy ($\beta = .264$, $q < .001$), placing the clearest observed user difference at the interpretation of functional breadth. Accounts of AI literacy emphasize users' capacity to recognize and evaluate AI [28], and prior work shows that AI knowledge shapes engagement with explanations [7]. Our result points to how prior knowledge helps users differentiate the functions represented in an assistant's output as a concrete question for further study. Evaluations of capability displays can therefore examine whose judgments change, as well as whether the average participant notices the manipulation.

### 6.2 When a rationale serves as evidence of intelligence

In the explanation-present condition, 75.4% of participants reported seeing the rationale, while its estimated effect on perceived intelligence remained near zero in both the factorial and structural analyses. Recognition alone leaves open how users evaluated the explanation. The study measured the availability of a brief rationale; its perceived usefulness and diagnostic value were unmeasured.

Miller treats explanation as a process shaped by an audience's needs [37], and Liao and colleagues show how those needs take the form of questions users want an AI system to answer [27]. These accounts suggest testing explanations as capability cues by examining whether a rationale helps users assess the reasoning behind a particular output. A future experiment could vary how directly the rationale addresses that task, measure its perceived diagnostic value and perceived usefulness, and examine the resulting intelligence judgment. An explanation intended to signal intelligence needs to give users a basis for judging the assistant's reasoning through the account it provides of a particular output. In this experiment, the rationale was generally noticed and its effect on that judgment remained near zero. This directs evaluation toward what users can infer about the assistant from the explanation they receive.

### 6.3 Capability judgments as material for personlike attribution

Perceived intelligence was positively associated with anthropomorphism under both MLR ($\beta$ = .393) and WLSMV ($\beta$ = .547). This relationship connects the capability judgments examined above to a different way of construing the assistant. Participants who regarded it as more intelligent also tended to attribute personlike qualities to it, even though anthropomorphic presentation was held constant across conditions.

The three-factor theory of anthropomorphism helps interpret this association by proposing that an agent's characteristics and behavior can activate knowledge that people use to understand it in human terms [8]. Judging an assistant as knowledgeable or cognitively capable may therefore provide material for personlike interpretation. The CASA tradition establishes more broadly that people can apply social expectations to computational systems [41]; the present finding adds specificity by showing that, even when anthropomorphic presentation was held constant, personlike construal covaried with perceived cognitive capability. Because both judgments were measured concurrently, the result does not establish that perceived intelligence caused anthropomorphism. It instead suggests that social attribution may be related not only to explicit humanlike design cues, but also to how users evaluate what an assistant can do.

The distinction becomes consequential when trust enters the analysis. Perceived intelligence and anthropomorphism remained empirically distinct, and their associations with trust took different forms. The resulting question is which aspect of trust accompanies a personlike interpretation once the assistant's perceived intelligence has been taken into account.

### 6.4 Personlike attribution concerns whose interests the assistant serves

Perceived intelligence had its strongest association with ability-based trust ($\beta$ = .651), followed by integrity ($\beta$ = .493) and benevolence ($\beta$ = .364). The Wald comparisons supported this ordering, showing that ability exceeded both integrity and benevolence, while integrity also exceeded benevolence ($q$ = .034). Anthropomorphism showed a different profile. Its additional association with benevolence ($\beta$ = .335) was larger than its associations with either integrity or ability (both $q$ < .001), which were individually nonsignificant. These comparisons establish differentiation across trust dimensions beyond a contrast between significant and nonsignificant paths.

The pattern sharpens what a social interpretation of AI contributes to a trust judgment. Ability-based trust concerns the assistant's capacity to do the work, so its close relationship with perceived intelligence is conceptually expected. Benevolence concerns whether the assistant appears to have the user's interests in mind [35]. Personlike attribution retained an association precisely with this orientation toward the user after perceived intelligence was included. In this setting, the social component of trust was most visible in judgments about whom the assistant seemed to serve.

Krämer and colleagues found a benevolence-specific result when manipulating a social interface cue [22]. Our study complements that finding by examining users' interpretations under a fixed anthropomorphic presentation, where variation in personlike attribution still had a selective relationship with benevolence. This also develops the distinction between cognitive and emotion-based trust discussed by Moussawi and colleagues [38]. Once perceived intelligence was accounted for, the additional trust associated with personlike attribution concerned whether the assistant appeared to care about the user's interests. This locates the distinctive contribution of anthropomorphism in perceived benevolence within a broader evaluation of the assistant's competence and trustworthiness.

This specificity changes the target for subsequent research. Studies of humanlike presentation and trust calibration [6, 20] can ask whether a cue changes confidence in task performance or the sense that an assistant acts in the user's interests. Keeping these outcomes separate makes it possible to identify a design that reassures users about the relationship while leaving their assessment of competence largely stable. The present findings supply an empirical basis for that distinction and a reason to retain it when evaluating socially expressive AI.

### 6.5 Evaluating AI as an interpreter of platform content

A platform-embedded assistant occupies a distinctive position by offering an interpretation of content that users can also encounter through the post and its surrounding discussion. Research on algorithmic awareness has examined how users infer computational processes from visible platform features [9]; work on AI-mediated communication examines how AI involvement bears on judgments within communication [12, 19]. Our study focuses on the assistant offering that interpretation. The trust distinction found here shows how confidence in its ability to analyze a post and a sense that it has the reader's interests in mind constitute separable evaluations of the same intermediary.

The controlled Reddit setting preserved the post, comments, scores, and other community signals while varying the assistant's functional display and rationale. Old Reddit provided the interface required by the extension. Its role in the study was to keep the surrounding platform evidence stable while examining judgments of the embedded assistant. Against this shared background, the trust profiles distinguish confidence in the AI's ability to interpret the post from a sense that it has the reader's interests in mind. Designing an assistant that appears beside a post thus calls for examining how readers assess its competence to interpret the content alongside its apparent orientation toward their interests.

The next step is to bring these sources of evidence into relation. Seering and colleagues situate social agents within community life [44]. Building on that perspective, experiments could vary whether visible community responses endorse or challenge an assistant's interpretation, then test which trust judgment changes. Agreement might support confidence in its competence, while a response about whose interests the assistant serves could bear more directly on benevolence. These are testable extensions of the differentiated trust pattern. They connect the evaluation of an embedded assistant to the social evidence available around it, giving platform research a way to study how users judge an AI interpreter within a wider discussion.

### 6.6 Limitations and future work

The experiment examined one controlled encounter with fixed, pre-generated outputs. Participants could observe the assistant but had limited opportunity to test its capabilities through interaction. Immediate self-reports capture initial evaluations; repeated tasks, behavioral reliance, and continued use would show how these judgments develop and guide action.

All focal perceptions were measured in the same post-exposure survey. Shared method variance, unmeasured general evaluations, and uncertain temporal order limit causal interpretation of the structural paths. Independently manipulating capability and anthropomorphic presentation, with measurements separated over time, would provide a stronger test of their respective roles in trust.

The multifunctional display combined a function-count label with topic tags, a summary, and a fact-check, producing more content than the single summary. Its effect therefore concerns this display package. A follow-up could hold content volume constant while varying functional breadth. The explanation result likewise concerns the availability of one brief rationale, with perceived explanation quality left for further measurement.

Finally, one COVID-19 vaccination post in Old Reddit and a U.S. Prolific sample limit transfer to other topics and contemporary platform layouts. Prior topic attitudes may matter, and anthropomorphism was relatively low ($M = 2.61$ on a seven-point scale). The literacy moderation was identified and estimated in the same comparatively high-scoring sample. Replication across a broader range of AI knowledge would establish how consistently that user difference appears.

## 7 Conclusion

Displaying additional functions reliably changed perceived functional breadth, while evidence for broader randomized changes was limited after correction. The structural model revealed how users' evaluations were differentiated: perceived intelligence was associated most strongly with ability-based trust and also with integrity and benevolence; anthropomorphism had an additional association concentrated in benevolence.

For platform-embedded AI, this pattern makes the social aspect of trust more specific. An assistant offers an interpretation of the content beside it, and users' judgments concern both its capacity to perform that work and its apparent orientation toward their interests. Distinguishing these dimensions also gives future research a more specific question: whether community feedback shapes judgments of an assistant's ability, integrity, and benevolence differently.

## References

[1] Yejin Bang, Samuel Cahyawijaya, Nayeon Lee, Wenliang Dai, Dan Su, Bryan Wilie, Holy Lovenia, Ziwei Ji, Tiezheng Yu, Willy Chung, Quyet V. Do, Yan Xu, and Pascale Fung. 2023. A Multitask, Multilingual, Multimodal Evaluation of ChatGPT on Reasoning, Hallucination, and Interactivity. arXiv:2302.04023 [cs]. doi:10.48550/arXiv.2302.04023

[2] Christoph Bartneck, Dana Kulić, Elizabeth Croft, and Susana Zoghbi. 2009. Measurement Instruments for the Anthropomorphism, Animacy, Likeability, Perceived Intelligence, and Perceived Safety of Robots. *International Journal of Social Robotics* 1, 1 (Jan. 2009), 71–81. doi:10.1007/s12369-008-0001-3

[3] Yoav Benjamini and Yosef Hochberg. 1995. Controlling the False Discovery Rate: A Practical and Powerful Approach to Multiple Testing. *Journal of the Royal Statistical Society Series B: Statistical Methodology* 57, 1 (Jan. 1995), 289–300. doi:10.1111/j.2517-6161.1995.tb02031.x

[4] Rocky Peng Chen, Echo Wen Wan, and Eric Levy. 2017. The effect of social exclusion on consumer preference for anthropomorphized brands. *Journal of Consumer Psychology* 27, 1 (2017), 23–34. doi:10.1016/j.jcps.2016.05.004

[5] Sungwoo Choi, Stephanie Q. Liu, and Anna S. Mattila. 2019. "How may i help you?" Says a robot: Examining language styles in the service encounter. *International Journal of Hospitality Management* 82 (Sept. 2019), 32–38. doi:10.1016/j.ijhm.2019.03.026

[6] Ewart J. de Visser, Samuel S. Monfort, Ryan McKendrick, Melissa A. B. Smith, Patrick E. McKnight, Frank Krueger, and Raja Parasuraman. 2016. Almost human: Anthropomorphism increases trust resilience in cognitive agents. *Journal of Experimental Psychology. Applied* 22, 3 (Sept. 2016),

331–349. doi:10.1037/xap0000092

[7] Upol Ehsan, Samir Passi, Q. Vera Liao, Larry Chan, I-Hsiang Lee, Michael Muller, and Mark O. Riedl. 2024. The who in XAI: How AI background shapes perceptions of AI explanations. In *Proceedings of the CHI conference on human factors in computing systems*. Association for Computing Machinery, 1–32. doi:10.1145/3613904.3642474

[8] Nicholas Epley, Adam Waytz, and John T. Cacioppo. 2007. On seeing human: A three-factor theory of anthropomorphism. *Psychological Review* 114, 4 (2007), 864–886. doi:10.1037/0033-295X.114.4.864

[9] Motahhare Eslami, Aimee Rickman, Kristen Vaccaro, Amirhossein Aleyasen, Andy Vuong, Karrie Karahalios, Kevin Hamilton, and Christian Sandvig. 2015. "I always assumed that I wasn't really that close to [her]": Reasoning about Invisible Algorithms in News Feeds. In *Proceedings of the 33rd annual ACM conference on human factors in computing systems*. Association for Computing Machinery, 153–162. doi:10.1145/2702123.2702556

[10] Xue Fan and Soyeon Kwon. 2026. Enhancing apology sincerity in AI bots: The role of anthropomorphic cues, perceived experience, and AI literacy. *Technological Forecasting and Social Change* 227 (June 2026), 124622. doi:10.1016/j.techfore.2026.124622

[11] Claes Fornell and David F. Larcker. 1981. Evaluating Structural Equation Models with Unobservable Variables and Measurement Error. *Journal of Marketing Research* 18, 1 (1981), 39–50. doi:10.2307/3151312

[12] Jarod Govers, Cherie Sew, Eduardo Velloso, Vassilis Kostakos, and Jorge Goncalves. 2026. Narratives and perspectives: How AI summaries steer users' opinions and engagement on social media. In *Proceedings of the 2026 CHI conference on human factors in computing systems*. Association for Computing Machinery, 1–19. doi:10.1145/3772318.3790945

[13] Shirley Gregor and Izak Benbasat. 1999. Explanations from Intelligent Systems: Theoretical Foundations and Implications for Practice. *MIS Quarterly* 23, 4 (1999), 497–530. doi:10.2307/249487

[14] Jeffrey T. Hancock, Mor Naaman, and Karen Levy. 2020. AI-mediated communication: Definition, research agenda, and ethical considerations. *Journal of Computer-Mediated Communication* 25, 1 (2020), 89–100. doi:10.1093/jcmc/zmz022

[15] Jörg Henseler, Christian M. Ringle, and Marko Sarstedt. 2015. A new criterion for assessing discriminant validity in variance-based structural equation modeling. *Journal of the Academy of Marketing Science* 43, 1 (2015), 115–135. doi:10.1007/s11747-014-0403-8

[16] Evelien Heyselaar. 2023. The CASA theory no longer applies to desktop computers. *Scientific Reports* 13, 1 (Nov. 2023), 19693. doi:10.1038/s41598-023-46527-9

[17] Shiyuan Huang, Siddarth Mamidanna, Shreedhar Jangam, Yilun Zhou, and Leilani H. Gilpin. 2023. Can Large Language Models Explain Themselves? A Study of LLM-Generated Self-Explanations. arXiv:2310.11207 [cs]. doi:10.48550/arXiv.2310.11207

[18] Andrew Hutchinson. 2025. X Expands Grok Translation on Posts In-Stream | Social Media Today. https://www.socialmediatoday.com/news/x-formerly-twitter-expands-grok-translations-posts/757620/

[19] Maurice Jakesch, Megan French, Xiao Ma, Jeffrey T. Hancock, and Mor Naaman. 2019. AI-mediated communication: How the perception that profile text was written by AI affects trustworthiness. In *Proceedings of the 2019 CHI conference on human factors in computing systems (Chi '19)*. Association for Computing Machinery, Glasgow, Scotland Uk, 1–13. doi:10.1145/3290605.3300469

[20] Theodore Jensen, Mohammad Maifi Hasan Khan, Md Abdullah Al Fahim, and Yusuf Albayram. 2021. Trust and anthropomorphism in tandem: The interrelated nature of automated agent appearance and reliability in trustworthiness perceptions. In *Proceedings of the 2021 ACM designing interactive systems conference (Dis '21)*. Association for Computing Machinery, Virtual Event, USA, 1470–1480. Number of pages: 11 tex.address: New York, NY, USA. doi:10.1145/3461778.3462102

[21] Alexandra D. Kaplan, Theresa T. Kessler, J. Christopher Brill, and P. A. Hancock. 2023. Trust in Artificial Intelligence: Meta-Analytic Findings. *Human Factors* 65, 2 (March 2023), 337–359. doi:10.1177/00187208211013988

[22] Nicole C. Krämer, Ivana Lamia, Hanne Siegert, Florian Wenda, and Lovis Suchmann. 2025. Tricking into Trusting? The Influence of Social Cues of a Generative AI on Perceived Trust. *ACM Trans. Interact. Intell. Syst.* 15, 4 (Dec. 2025), 27:1–27:17. doi:10.1145/3771844

[23] Anastasia Kuzminykh, Jenny Sun, Nivetha Govindaraju, Jeff Avery, and Edward Lank. 2020. Genie in the Bottle: Anthropomorphized Perceptions of Conversational Agents. In *Proceedings of the 2020 CHI Conference on Human Factors in Computing Systems (CHI '20)*. Association for Computing Machinery, New York, NY, USA, 1–13. doi:10.1145/3313831.3376665

[24] John D. Lee and Katrina A. See. 2004. Trust in automation: Designing for appropriate reliance. *Human Factors* 46, 1 (2004), 50–80. doi:10.1518/hfes.46.1.50_30392

[25] Won-jun Lee and Seungjae Shin. 2018. Effects of Product Smartness on Satisfaction: Focused on the Perceived Characteristics of Smartphones. *Journal of Theoretical and Applied Electronic Commerce Research* 13, 2 (May 2018), 1–14. doi:10.4067/S0718-18762018000200102

[26] Qingchuan Li, Yan Luximon, and Jiaxin Zhang. 2023. The Influence of Anthropomorphic Cues on Patients' Perceived Anthropomorphism, Social Presence, Trust Building, and Acceptance of Health Care Conversational Agents: Within-Subject Web-Based Experiment. *Journal of Medical Internet Research* 25 (Aug. 2023), e44479. doi:10.2196/44479

[27] Q. Vera Liao, Daniel Gruen, and Sarah Miller. 2020. Questioning the AI: Informing Design Practices for Explainable AI User Experiences. In *Proceedings of the 2020 CHI conference on human factors in computing systems*. Association for Computing Machinery, 1–15. doi:10.1145/3313831.3376590

[28] Duri Long and Brian Magerko. 2020. What is AI literacy? Competencies and design considerations. In *Proceedings of the 2020 CHI conference on human factors in computing systems*. Association for Computing Machinery, 1–16. doi:10.1145/3313831.3376727

[29] Ewa Luger and Abigail Sellen. 2016. "Like having a really bad PA": The gulf between user expectation and experience of conversational agents. In *Proceedings of the 2016 CHI conference on human factors in computing systems*. Association for Computing Machinery, 5286–5297. doi:10.1145/2858036.2858288

[30] Ning Ma, Ruslana Khynevych, Yunqiang Hao, and Yahui Wang. 2025. Effect of anthropomorphism and perceived intelligence in chatbot avatars of visual design on user experience: accounting for perceived empathy and trust. *Frontiers in Computer Science* 7 (May 2025). doi:10.3389/fcomp.2025.1531976

[31] Markus Makkonen, Markus Salo, and Henri Pirkkalainen. 2022. What Makes a (Ro)bot Smart? Examining the Antecedents of Perceived Intelligence in the Context of Using Physical Robots, Software Robots, and Chatbots at Work. *MCIS 2022 Proceedings* (Oct. 2022). https://aisel.aisnet.org/mcis2022/3

[32] Chenchen Mao, Hanjing Shi, Haiyan Jia, Daniel Unhuryan, Eric Baumer, and Dominic DiFranzo. 2026. Outer Limits: An Experimental Approach to Controlled Content Manipulation within the Reddit Interface. arXiv:2608.10115 [cs.HC]. doi:10.48550/arXiv.2608.10115

[33] Chenchen Mao, Hanjing Shi, Haiyan Jia, Emily Wegrzyn, and Dominic DiFranzo. 2026. When Readability and Source Retention Diverge: An Evaluability Gap in AI Translation. arXiv:2608.19083 [cs.HC]. doi:10.48550/arXiv.2608.19083

[34] Nestor Maslej, Loredana Fattorini, Raymond Perrault, Yolanda Gil, Vanessa Parli, Njenga Kariuki, Emily Capstick, Anka Reuel, Erik Brynjolfsson, John Etchemendy, Katrina Ligett, Terah Lyons, James Manyika, Juan Carlos Niebles, Yoav Shoham, Russell Wald, Toby Walsh, Armin Hamrah, Lapo Santarlasci, Julia Betts Lotufo, Alexandra Rome, Andrew Shi, and Sukrut Oak. 2025. Artificial Intelligence Index Report 2025. arXiv:2504.07139 [cs.AI]. doi:10.48550/arXiv.2504.07139

[35] Roger C. Mayer, James H. Davis, and F. David Schoorman. 1995. An Integrative Model of Organizational Trust. *The Academy of Management Review* 20, 3 (1995), 709–734. doi:10.2307/258792

[36] Ivan Mehta. 2025. X now lets you query Grok by mentioning it in replies. https://techcrunch.com/2025/03/07/x-now-lets-you-query-grok-by-mentioning-it-in-replies/

[37] Tim Miller. 2019. Explanation in artificial intelligence: Insights from the social sciences. *Artificial Intelligence* 267 (Feb. 2019), 1–38. doi:10.1016/j.artint.2018.07.007

[38] Sara Moussawi and Raquel Benbunan-Fich. 2021. The effect of voice and humour on users' perceptions of personal intelligent agents. *Behaviour & Information Technology* 40, 15 (Nov. 2021), 1603–1626. doi:10.1080/0144929X.2020.1772368

[39] Sara Moussawi and Marios Koufaris. 2019. Perceived intelligence and perceived anthropomorphism of personal intelligent agents: Scale development and validation. *Proceedings of the 52nd Hawaii International Conference on System Sciences, 115–124* (Jan. 2019). doi:10.24251/hicss.2019.015

[40] Sara Moussawi, Marios Koufaris, and Raquel Benbunan-Fich. 2021. How perceptions of intelligence and anthropomorphism affect adoption of personal intelligent agents. *Electronic Markets* 31, 2 (June 2021), 343–364. doi:10.1007/s12525-020-00411-w

[41] Clifford Nass and Youngme Moon. 2000. Machines and mindlessness: Social responses to computers. *Journal of Social Issues* 56, 1 (2000), 81–103. doi:10.1111/0022-4537.00153

[42] Mijke Rhemtulla, Patricia É. Brosseau-Liard, and Victoria Savalei. 2012. When can categorical variables be treated as continuous? A comparison of robust continuous and categorical SEM estimation methods under suboptimal conditions. *Psychological Methods* 17, 3 (2012), 354–373. doi:10.1037/a0029315

[43] Yves Rosseel. 2012. lavaan: An R Package for Structural Equation Modeling. *Journal of Statistical Software* 48 (May 2012), 1–36. doi:10.18637/jss.v048.i02

[44] Joseph Seering, Michal Luria, Connie Ye, Geoff Kaufman, and Jessica Hammer. 2020. It takes a village: Integrating an adaptive chatbot into an online gaming community. In *Proceedings of the 2020 CHI conference on human factors in computing systems*. Association for Computing Machinery, 1–13. doi:10.1145/3313831.3376708

[45] Christianna Silva. 2025. Meta wants AI to write your Instagram comments. https://mashable.com/article/meta-test-instagram-ai-caption-comments

[46] S Shyam Sundar. 2020. Rise of Machine Agency: A Framework for Studying the Psychology of Human–AI Interaction (HAII). *Journal of Computer-Mediated Communication* 25, 1 (Jan. 2020), 74–88. doi:10.1093/jcmc/zmz026

[47] Isaac Triguero, Daniel Molina, Javier Poyatos, Javier Del Ser, and Francisco Herrera. 2024. General Purpose Artificial Intelligence Systems (GPAIS): Properties, definition, taxonomy, societal implications and responsible governance. *Information Fusion* 103 (March 2024), 102135. doi:10.1016/j.inffus.2023.102135

[48] Stephanie M. Tully, Chiara Longoni, and Gil Appel. 2025. Lower Artificial Intelligence Literacy Predicts Greater AI Receptivity. *Journal of Marketing* 89, 5 (Sept. 2025), 1–20. doi:10.1177/00222429251314491

[49] Qiaosi Wang, Koustuv Saha, Eric Gregori, David Joyner, and Ashok Goel. 2021. Towards Mutual Theory of Mind in Human-AI Interaction: How Language Reflects What Students Perceive About a Virtual Teaching Assistant. In *Proceedings of the 2021 CHI Conference on Human Factors in Computing Systems (CHI '21)*. Association for Computing Machinery, New York, NY, USA, 1–14. doi:10.1145/3411764.3445645

[50] Weiquan Wang and Izak Benbasat. 2007. Recommendation Agents for Electronic Commerce: Effects of Explanation Facilities on Trusting Beliefs. *Journal of Management Information Systems* 23, 4 (May 2007), 217–246. doi:10.2753/MIS0742-1222230410

[51] Geert Wood, Elena Nuñez Castellar, and Wijnand IJsselsteijn. 2025. An Exploratory Study Into the Impact of AI Literacy Training on Anthropomorphism and Trust in Conversational AI. In *Artificial Intelligence in HCI*, Helmut Degen and Stavroula Ntoa (Eds.). Springer Nature Switzerland, Cham, 301–322. doi:10.1007/978-3-031-93415-5_18

[52] Ming Yin, Jennifer Wortman Vaughan, and Hanna Wallach. 2019. Understanding the effect of accuracy on trust in machine learning models. In *Proceedings of the 2019 CHI conference on human factors in computing systems*. Association for Computing Machinery, 1–12. doi:10.1145/3290605.3300509

[53] Jing Zang and Myounghoon Jeon. 2022. The Effects of Transparency and Reliability of In-Vehicle Intelligent Agents on Driver Perception, Takeover Performance, Workload and Situation Awareness in Conditionally Automated Vehicles. *Multimodal Technologies and Interaction* 6, 9 (Sept. 2022). doi:10.3390/mti6090082

[54] Tingru Zhang, Weitao Li, Weixing Huang, and Liang Ma. 2024. Critical roles of explainability in shaping perception, trust, and acceptance of autonomous vehicles. *International Journal of Industrial Ergonomics* 100 (March 2024), 103568. doi:10.1016/j.ergon.2024.103568

## A Appendix for the measures used in the Study

This appendix reproduces the item wording for the six constructs in the retained SEM and for objective AI literacy. The wording below reflects the items presented to participants.

### A.1 Perceived Multifunctionality

Indicate the extent to which you agree with each of the following statements:

- This AI assistant has multiple functions.
- This AI assistant can do a lot.
- This AI assistant performs multiple tasks.
- This AI assistant fulfills multiple functional needs.

**Answer options:** 1 = Strongly disagree, to 7 = Strongly agree.

### A.2 Perceived Intelligence

Please rate your impression of the AI assistant on each of the following scales:

- Incompetent (1) to Competent (5);
- Ignorant (1) to Knowledgeable (5);
- Irresponsible (1) to Responsible (5);
- Unintelligent (1) to Intelligent (5);
- Foolish (1) to Sensible (5).

**Answer options:** Five response positions ranging from 1 (the adjective shown on the left) to 5 (the adjective shown on the right).

### A.3 Anthropomorphism

Indicate the extent to which you agree with each of the following statements:

- The AI assistant feels like a person.
- I think of the AI assistant as a person.
- The AI assistant has its own personality.
- The AI assistant has its own intentions.

**Answer options:** 1 = Strongly disagree, to 7 = Strongly agree.

### A.4 Trust Dimensions

Indicate the extent to which you agree with each of the following statements:

#### *A.4.1 Benevolence.*

- This AI assistant has my best interests in mind.
- This AI assistant is concerned with helping me.
- This AI assistant is attentive to what would be helpful for me.

#### *A.4.2 Integrity.*

- This AI assistant provides unbiased information.

- This AI assistant is honest.
- I consider this AI assistant to have integrity.

#### *A.4.3 Ability.*

- This AI assistant is highly capable of helping users understand the Reddit post.
- This AI assistant has the expertise needed to analyze the Reddit post.
- This AI assistant has the ability to identify important information in the Reddit post.
- This AI assistant has good knowledge about the content it is analyzing.
- This AI assistant is capable of providing helpful information about the Reddit post.

**Answer options:** 1 = Strongly disagree, to 7 = Strongly agree.

### A.5 Objective AI Literacy

Objective AI literacy was scored automatically in Qualtrics. Each correct response contributed one point, producing a score from 0 to 17. The administered items and response options were:

- **Which of the following is NOT powered by AI?** Self-driving cars; Google's search algorithm; a basic calculator; chatbots. *Correct: a basic calculator.*
- **Which form of intelligence involves emotional understanding and social skills?** Machine intelligence; human intelligence; animal intelligence; artificial general intelligence. *Correct: human intelligence.*
- **Which of the following fields contributes to the development of artificial intelligence?** Computer science; mathematics; psychology; all of the above. *Correct: all of the above.*
- **What is the name of AI systems that can perform any intellectual task that a human can?** Narrow AI; general AI; weak AI; strong AI. *Correct: general AI.*
- **In which area does AI typically excel?** Emotional understanding; pattern recognition; moral reasoning; creativity. *Correct: pattern recognition.*
- **Which of the following is NOT a likely future application of AI?** Personalized healthcare; emotional robots; time travel; sustainable energy management. *Correct: time travel.*
- **What is a common form of knowledge representation in AI?** Neural networks; waterfall model; agile methodology; SWOT analysis. *Correct: neural networks.*
- **Which algorithmic approach is commonly used for decision-making in AI?** Dijkstra's algorithm; depth-first search; decision trees; Fourier transform. *Correct: decision trees.*
- **What is the first step in a typical machine learning process?** Data collection; model selection; prediction; model evaluation. *Correct: data collection.*
- **Who is primarily responsible for an AI system's ethical considerations?** The AI system itself; data providers; human developers; end-users. *Correct: human developers.*
- **Which of the following is an example of metadata?** A spreadsheet of numbers; column headers in a table; a chart visualization; raw sensor data. *Correct: column headers in a table.*
- **How do supervised machine-learning algorithms learn?** From labeled data; from rewards and punishments; by observing human behavior; from intrinsic motivation. *Correct: from labeled data.*
- **Why should data not be taken at face value?** It is always inaccurate; it requires interpretation; it is self-explanatory; it is always biased. *Correct: it requires interpretation.*

- **How can an AI system interact with the physical world?** By planning movements; by reacting to sensor inputs; by actuating motors; all of the above. *Correct: all of the above.*
- **Which of the following sensors enable an AI system to perceive the world?** Cameras; microphones; thermometers; all of the above. *Correct: all of the above.*
- **What is a key ethical issue surrounding AI?** Algorithmic efficiency; CPU usage; privacy; code readability. *Correct: privacy.*
- **Which statement best describes the programmability of AI systems?** They cannot be programmed by humans; they program themselves; they are programmed using data; they are programmed by computer code. *Correct: they are programmed by computer code.*

## B Full discriminant-validity results for the retained six-factor model

Table 4 reports every pair from the retained six-factor CFA. Fornell–Larcker requires the absolute latent correlation to remain below the smaller square root of AVE. HTMT values below .90 satisfy the second criterion used in the paper.

| **Construct pair** | **Latent $r$** | **Minimum $\sqrt{\text{AVE}}$** | **HTMT** |
|---|---|---|---|
| Perceived intelligence–Perceived multifunctionality | .584 | .864 | .585 |
| Anthropomorphism–Perceived multifunctionality | .434 | .859 | .441 |
| Anthropomorphism–Perceived intelligence | .382 | .859 | .370 |
| Benevolence–Perceived multifunctionality | .400 | .883 | .401 |
| Benevolence–Perceived intelligence | .491 | .864 | .496 |
| Benevolence–Anthropomorphism | .479 | .859 | .490 |
| Integrity–Perceived multifunctionality | .359 | .849 | .362 |
| Integrity–Perceived intelligence | .517 | .849 | .553 |
| Integrity–Anthropomorphism | .265 | .849 | .248 |
| Integrity–Benevolence | .617 | .849 | .636 |
| Ability–Perceived multifunctionality | .483 | .891 | .487 |
| Ability–Perceived intelligence | .650 | .864 | .650 |
| Ability–Anthropomorphism | .271 | .859 | .269 |
| Ability–Benevolence | .514 | .883 | .514 |
| Ability–Integrity | .670 | .849 | .676 |

Table 4. Latent correlations, the smaller square root of AVE for each pair, and HTMT values for the retained six-factor CFA. All pairs pass both criteria.

## C Exploratory Objective AI Literacy Moderation Screen

We examined objective AI literacy as a potential moderator of each of the 10 structural relationships represented in the hypothesized SEM. For this exploratory screening stage, each interaction was estimated in a separate OLS regression using standardized composite scores and HC3 robust standard errors. The resulting $p$ values were adjusted using the Benjamini–Hochberg procedure across the 10 interaction tests.

As shown in Table 5, only the interaction between assigned multifunctionality and objective AI literacy in predicting perceived multifunctionality remained statistically significant after correction ($q < .001$). This interaction was subsequently incorporated into and re-estimated within the latent structural model. The remaining interactions were not included because they did not survive the multiplicity correction.

| **Relationship moderated by objective AI literacy** | $b$ | **HC3** $SE$ | **95% CI** | $p$ | $q$ |
|---|---|---|---|---|---|
| Assigned multifunctionality → Perceived multifunctionality | .520 | .129 | [.267, .772] | < .001 | < .001*** |
| Assigned explanation → Perceived intelligence | .114 | .095 | [−.072, .300] | .231 | .586 |
| Perceived multifunctionality → Perceived intelligence | −.067 | .055 | [−.174, .040] | .219 | .586 |
| Perceived intelligence → Anthropomorphism | −.002 | .042 | [−.084, .081] | .970 | .970 |
| Perceived intelligence → Benevolence | −.018 | .044 | [−.105, .068] | .677 | .946 |
| Anthropomorphism → Benevolence | .021 | .043 | [−.064, .106] | .634 | .946 |
| Perceived intelligence → Integrity | −.027 | .074 | [−.172, .118] | .718 | .946 |
| Anthropomorphism → Integrity | −.055 | .046 | [−.145, .035] | .234 | .586 |
| Perceived intelligence → Ability | −.019 | .102 | [−.218, .180] | .852 | .946 |
| Anthropomorphism → Ability | −.010 | .043 | [−.094, .075] | .826 | .946 |

Table 5. Exploratory screening of objective AI literacy as a moderator of the 10 structural relationships represented in the hypothesized SEM. Each interaction was estimated in a separate OLS regression using standardized composite scores and HC3 robust standard errors. The $q$ values were calculated using the Benjamini–Hochberg procedure across the 10 interaction tests. $^{***}q < .001$.

## D Full Experimental Interface

Figure 4 provides an expanded view of the experimental interface. Outer Limits opened the same underlying Reddit URL for all participants and locally presented preconstructed study materials for the post title, author, posting time, post content, post score, commenters, comments, comment timestamps, and comment scores. These elements were held constant across conditions. Only assigned multifunctionality and assigned explanation varied.

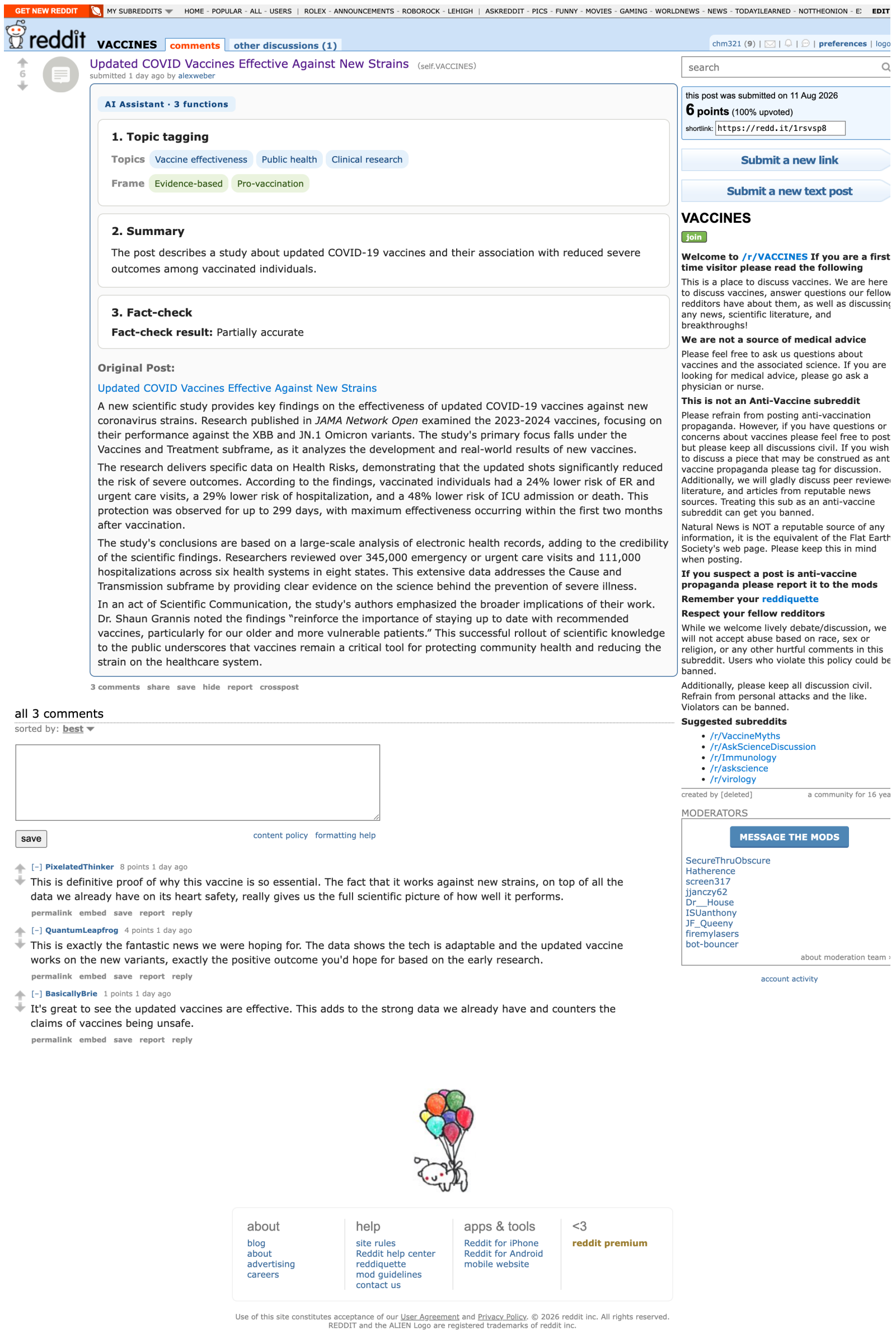

Fig. 4. **Full experimental interface in the multifunctional/explanation-absent condition.** The AI assistant provides three functions: topic tagging, a summary, and a fact-check. The explanation panel is absent in this condition. The post, engagement indicators, and comments were prepared in advance and rendered locally. Participant actions remained within the study interface.